\documentclass{article} %

\usepackage[nonatbib,final]{neurips_2019}
\usepackage{fancyhdr}
\usepackage[utf8]{inputenc}    
\usepackage[T1]{fontenc}

\usepackage{amsfonts}
\usepackage{amssymb}
\usepackage{amsthm}
\usepackage{amsmath}
\usepackage{caption}
\usepackage[inline]{enumitem}
\setlist{itemsep=0em, topsep=0em, parsep=0em}
\setlist[enumerate]{label=(\alph*)}
\usepackage{etoolbox}
\usepackage{stmaryrd}
\usepackage[dvipsnames]{xcolor}
\usepackage{ifthen}

\usepackage{microtype}

\usepackage{nicefrac}
\usepackage{bm}
\usepackage{physics}
\usepackage{tipa} %
\usepackage{mathtools} %
\usepackage{color}
\usepackage{wrapfig}
\usepackage{tikz}
\usetikzlibrary{
	cd,
	math,
	decorations.markings,
	decorations.pathreplacing,
	positioning,
	arrows.meta,
	circuits.logic.US,
	shapes,
	calc,
	fit,
	trees,
	quotes,
        positioning,intersections
}
\usepackage[framemethod=tikz]{mdframed}

\newcommand*{\figref}[1]{\figurename~\ref{#1}}

\usepackage[backend=biber,style=numeric-comp,sorting=none,eprint=false,doi=false,isbn=false,url=false,date=short,maxcitenames=2]{biblatex}
\usepackage[colorlinks,linkcolor=darkblue,citecolor=darkblue,urlcolor=darkblue,breaklinks=true]{hyperref}
\usepackage{url}

\usepackage{booktabs}

\bibliography{library}

\newcommand*{\citep}[1]{\parencite{#1}}
\newcommand*{\citet}[1]{\textcite{#1}}

\usepackage[capitalize]{cleveref} %

\usepackage{graphicx}

\newcommand{\iso}{\cong}

\newcommand{\Cat}[1]{\mathbf{#1}}

\DeclareMathOperator{\Ca}{\mathcal{C}}
\DeclareMathOperator{\Da}{\mathcal{D}}

\DeclareMathOperator{\id}{\mathsf{id}}

\makeatletter
\providecommand*{\xmapstofill@}{%
  \arrowfill@{\mapstochar\relbar}\relbar\rightarrow
}
\providecommand*{\xmapsto}[2][]{%
  \ext@arrow 0395\xmapstofill@{#1}{#2}%
}

\makeatletter
\def\slashedarrowfill@#1#2#3#4#5{%
  $\m@th\thickmuskip0mu\medmuskip\thickmuskip\thinmuskip\thickmuskip
   \relax#5#1\mkern-7mu%
   \cleaders\hbox{$#5\mkern-2mu#2\mkern-2mu$}\hfill
   \mathclap{#3}\mathclap{#2}%
   \cleaders\hbox{$#5\mkern-2mu#2\mkern-2mu$}\hfill
   \mkern-7mu#4$%
}
\def\rightslashedarrowfill@{%
  \slashedarrowfill@\relbar\relbar\mapstochar\rightarrow}
\newcommand\xslashedrightarrow[2][]{%
  \ext@arrow 0055{\rightslashedarrowfill@}{#1}{#2}}
\makeatother

\theoremstyle{definition}
\newtheorem{defn}{Definition}

\newtheorem{thm}[defn]{Theorem}

\newtheorem{prop}[defn]{Proposition}

\definecolor{darkblue}{rgb}{0,0,0.7}

\usepackage{img/tikzit}

\tikzstyle{white dot}=[fill=white, draw=black, shape=circle]
\tikzstyle{medium box}=[fill=white, draw=black, shape=rectangle, minimum width=1.5cm, minimum height=0.66cm]
\tikzstyle{rect}=[fill=white, draw=black, shape=rectangle]

\tikzstyle{pointy1}=[->]
\tikzstyle{dashed}=[-, dashed]
\tikzstyle{dash-pointy}=[->, dashed]

\title{Radically Compositional Cognitive Concepts}
\author{%
  Toby B. St Clere Smithe\thanks{\url{http://tsmithe.net}} \\
  Department of Experimental Psychology\\
  University of Oxford \\
  Oxford, UK \\
  OX2 6NW \\
  \texttt{toby.smithe@psy.ox.ac.uk}
}

\begin{document}
\maketitle

\begin{abstract}
  Despite ample evidence that our concepts, our cognitive architecture, and
  mathematics itself are all deeply compositional, few models take advantage of
  this structure. We therefore propose a radically compositional approach to
  computational neuroscience, drawing on the methods of applied category
  theory. We describe how these tools grant us a means to overcome complexity
  and improve interpretability, and supply a rigorous common language for
  scientific modelling, analogous to the type theories of computer science. As a
  case study, we sketch how to translate from compositional narrative concepts
  to neural circuits and back again.
\end{abstract}

\section{Introduction}

How should we conceive of concepts, of cognition, or of circuit computation? We
argue: compositionally. Composition is the tool by which we construct complex
concepts: constantly, informally, and automatically. But it is not just our
concepts that are compositional: it has long been noted that our cognitive
architecture is modular \citep{fodor1983modularity,marcus2018deep}, and much of
cognitive neuroscience relies on the well-tested assumption that this modularity
maps onto the structure of the brain. Mathematics itself is increasingly
recognized as compositional \citep{kock2006synthetic,bauer2017hott}, and strong
compositionality is increasingly used in software engineering to improve
correctness and code reusability
\citep{fischer2017haskell,haftmann2010higher}. Despite this simplifying power,
few models in cognitive science and machine learning are actually compositional,
even when modelling compositionality itself (\emph{eg.},
\citep{chang2016compositional,piantadosi2016logical}).

We propose instead taking compositionality seriously, using the mathematics of
composition---category theory---and show how doing so allows us to translate
concepts between contexts and levels: from abstract concepts themselves
\citep{phillips2010categorial,bolt2019interacting} to their possible realization
in a circuit model. This paper is an abridged version of a work in progress,
provisionally to appear in \emph{Compositionality}, and we defer many formal
details and proofs to that manuscript. In §\ref{sec:math} we introduce the
background mathematics. In §\ref{sec:concepts}, we introduce the conceptual
setting. §\ref{sec:circ} shows how to translate concepts to circuits, and
§\ref{sec:back} suggests how to translate back again.

\section{Category theory and the `structure of structure'} \label{sec:math}

A category is a very simple structure, capturing only what is necessary to
enforce compositionality:

\defn{A \textbf{category} $\Ca$ is a set of objects $\Ca_0$ such that for any
  two objects $x, y \in \Ca_0$ there is a set $\Ca(x, y)$ of arrows ${x \to y}$
  obeying a \textbf{composition rule}: for any arrow $f : x \to y$ and any arrow
  $g : y \to z$, there is a composite arrow $g \circ f : x \to z$. Every object
  $x \in \Ca_0$ has an \textbf{identity} arrow, $\id_x \in \Ca(x, x)$.}

Being so general, almost all concepts can be formalized categorically. For
example, there is a category whose objects are parts of speech and whose arrows
are grammatical relations; and a category of vector spaces and linear maps.
Category theory can also be applied to itself: 
there is a
category $\Cat{Cat}$ of categories, whose arrows are called functors:

\defn{A \textbf{functor} is a structure-preserving arrow $F : \Ca \to \Da$
  between categories, mapping identities to identities and respecting
  composition: $F(\id_x) = \id_{F(x)}$, and $F(g \circ f) = F(g) \circ F(f)$.}

Functors permit translating concepts from one category to another. A classic
example is \emph{functorial semantics} for language: one constructs a functor
from a syntactic category (modelling grammar) to a semantic category (modelling
meaning) that witnesses the compositionality of linguistic content.
There is
frequently also a notion of parallel composition, which we denote
$\otimes$. This situation is captured by the notion of monoidal category:

\defn{A category $\Ca$ is \textbf{monoidal} if it is equipped with a functor
  $\otimes : \Ca \times \Ca \to \Ca$ such that objects $x$ and $y$ and arrows $f
  : x \to a$ and $g : y \to b$ can be paired into compound objects $x \otimes y$
  and `parallel' arrows $f \otimes g : x \otimes y \to a \otimes b$. The functor
  $\otimes$ must satisfy some canonical coherence conditions, such as
  associativity, which are omitted for brevity. Many monoidal categories are
  \textbf{symmetric}, meaning that $a \otimes b \iso b \otimes a$.}

Monoidal categories admit a calculus of \emph{string diagrams}, in which arrows
are represented by boxes on wires. Wires are annotated by their object type. %
Sequential composition is achieved by
connecting wires of the same type, and parallel composition by parallel
placement; in a symmetric monoidal category, wires can cross and be uncrossed.
In the next section, we will meet some examples; for more details, see
\citep{baez2010physics}. Meanwhile, we emphasize that any network or process
diagram or computational graph is almost surely an arrow in some monoidal
category.

\textbf{Categorical probability theory.}
We review the monoidal categories of measurable spaces and of Bayesian
networks. For details, we refer the reader to \citep{panangaden1999category}.

\prop{There is a category $\Cat{Meas}$ whose objects are measurable spaces $(X,
  \Sigma_X)$ and whose arrows are measurable functions. Its monoidal product is
  the Cartesian product $\times$ of sets.}

\defn{The \emph{Giry monad} $G : \Cat{Meas} \to \Cat{Meas}$ takes a measurable
  space $(X, \Sigma_X)$ to the measurable space of (sub)probability
  distributions ${\Sigma_X \to [0, 1]}$ over $X$.
  }

\defn{The category $\Cat{BayesNets}$ is the \emph{Kleisli category} of $G$. Its
  objects are again measurable spaces. Arrows $X \to Y$ in $\Cat{BayesNets}$ are
  arrows $X \to G(Y)$ in $\Cat{Meas}$. The identity map $X \to G(X)$ takes a
  point in $X$ to the Dirac delta distribution over that point.
}

Since $G(Y)$ is equivalently the space of arrows $\{\Sigma_Y \to [0, 1]\}$, we
can think of an arrow $X \to G(Y)$ as equivalently an arrow $X \to \{\Sigma_Y
\to [0, 1]\}$. As in functional programming, we can `uncurry' this to an arrow
$X \times \Sigma_Y \to [0, 1]$. This is a \emph{Markov kernel}, the
measure-theoretic form of a conditional probability distribution $p(y | x)$, and
the composite $X \to G(Z)$ of $X \to G(Y)$ and $Y \to G(Z)$ is $\int_y p(z | y)
\, p(y | x) \, dy = \int_y p(z, y | x) \, dy = p(z | x)$. Henceforth, we will
often write just $y|x$ for an arrow $X \to G(Y)$ in $\Cat{BayesNets}$. An arrow
$1 \to G(X)$ from the unit space $1 = \{*\}$ is a plain (unconditional)
distribution over $X$.

We leave the monoidal structure $\otimes$ of $\Cat{BayesNets}$ informal, as the
intuition is familiar: it takes two distributions to their corresponding joint
distribution. Consequently, we can think of any arrow in $\Cat{BayesNets}$ as
describing a Bayesian network.

\section{Compositionality of narrative concepts} \label{sec:concepts}
\begin{figure}
  \centering
  \includegraphics[height=4cm]{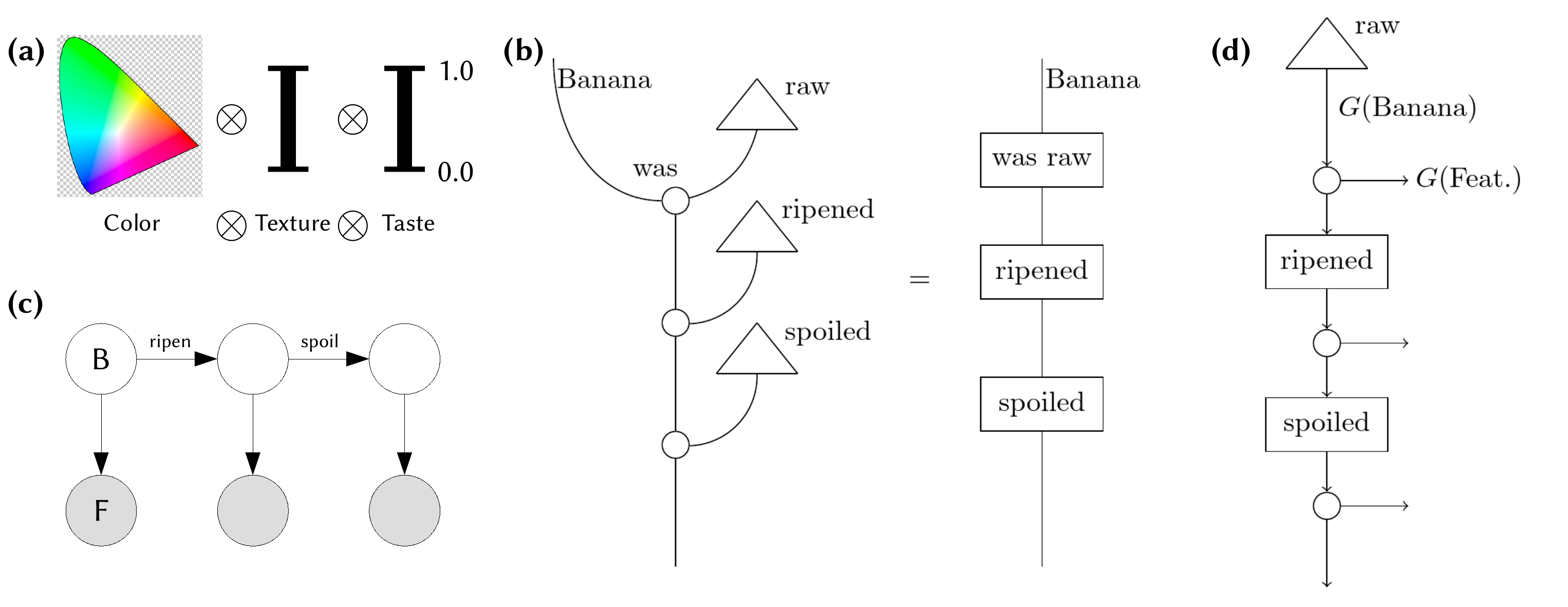}
  \caption{\label{fig:banana} We model a simple narrative about a banana, as it
    transitions from raw to spoiled. \textbf{(a)} The feature-level (F)
    representation of the high-level `banana' (B) concept is the convex monoidal
    product of color, texture (1.0 = `soft') and taste (1.0 =
    `sweet'). \textbf{(b)} The DisCoCirc representation of the ripening
    banana. \textbf{(c)} Traditional state-space model of (b). \textbf{(d)}
    `String diagram' representation of (c).}
\end{figure}

We choose as our model of concept representation Gärdenfors' classic `conceptual
spaces' \citep{gardenfors2004conceptual}, and as our model of concept
compositionality the `DisCoCat' model of Coecke et al
\cite{coecke2010mathematical,bolt2019interacting}. In this account, complex
concepts are composed from simple ones according to a grammar whose rules are
witnessed by the arrows of a syntactic category, just as described in
§\ref{sec:math} for the functorial semantics of language. Crucially, concepts
are modelled semantically as living in convex spaces
\citep{bolt2019interacting}, as illustrated in \figref{fig:banana}(a). This
point of view has recently received neurological validation
\citep{bellmund2018navigating}.

However, this basic account lacks two features: narrative (\emph{i.e.},
dynamics); and any concrete realization. To supply narrative, we can model
sentences as inducing semantic transitions (`DisCoCirc',
\citep{coecke2019mathematics}) in the underlying convex spaces -- a view which
accords with evidence from navigation in hippocampal circuits
\citep{bellmund2018navigating,behrens2018cognitive}. Next, we note that
distributions in $\Cat{BayesNets}$ form convex spaces, and that updating these
states preserves this convex structure \citep{cho2015introduction}. Thus we can
model narrative as a `conceptual' state-space model
(\figref{fig:banana}(b)-(d)). Next, we map this onto a prototypical neural
circuit.

\section{From compositional concepts to compositional circuits ...} \label{sec:circ}
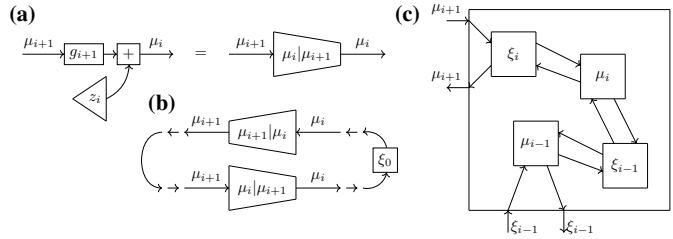
\begin{figure}
  \centering
  \resizebox{9cm}{!}{
    \begin{tikzpicture}
	\begin{pgfonlayer}{nodelayer}
		\node [style=rect] (0) at (-12.25, 4) {$g_{i+1}$};
		\node [style=none] (1) at (-15, 4) {};
		\node [style=rect] (2) at (-10.25, 4) {$+$};
		\node [style=none] (3) at (-12.75, 2) {};
		\node [style=none] (4) at (-11.25, 3) {};
		\node [style=none] (5) at (-11.25, 1) {};
		\node [style=none] (6) at (-11.75, 2) {$z_i$};
		\node [style=none] (7) at (-11.25, 2) {};
		\node [style=none] (8) at (-8.25, 4) {};
		\node [style=none] (9) at (-14.25, 4.5) {};
		\node [style=none] (10) at (-14.25, 4.5) {$\mu_{i+1}$};
		\node [style=none] (11) at (-9, 4.5) {$\mu_{i}$};
		\node [style=none] (12) at (-5.75, 4) {};
		\node [style=none] (13) at (-3.75, 4) {};
		\node [style=none] (14) at (-3.75, 5) {};
		\node [style=none] (15) at (-3.75, 3) {};
		\node [style=none] (16) at (-0.75, 3.5) {};
		\node [style=none] (17) at (-0.75, 4.5) {};
		\node [style=none] (18) at (1.25, 4) {};
		\node [style=none] (19) at (-0.75, 4) {};
		\node [style=none] (20) at (-4.75, 4.5) {$\mu_{i+1}$};
		\node [style=none] (21) at (0.25, 4.5) {$\mu_{i}$};
		\node [style=none] (22) at (-2.25, 4) {$\mu_i | \mu_{i+1}$};
		\node [style=none] (23) at (-7, 4) {$=$};
		\node [style=none] (24) at (-7.75, -2) {};
		\node [style=none] (25) at (-5.75, -2) {};
		\node [style=none] (26) at (-5.75, -1) {};
		\node [style=none] (27) at (-5.75, -3) {};
		\node [style=none] (28) at (-2.75, -2.5) {};
		\node [style=none] (29) at (-2.75, -1.5) {};
		\node [style=none] (30) at (-0.75, -2) {};
		\node [style=none] (31) at (-2.75, -2) {};
		\node [style=none] (32) at (-6.75, -1.5) {$\mu_{i+1}$};
		\node [style=none] (33) at (-1.75, -1.5) {$\mu_i$};
		\node [style=none] (34) at (-4.25, -2) {$\mu_i | \mu_{i+1}$};
		\node [style=none] (35) at (-0.75, 0.5) {};
		\node [style=none] (36) at (-2.75, 0.5) {};
		\node [style=none] (37) at (-2.75, 1.5) {};
		\node [style=none] (38) at (-2.75, -0.5) {};
		\node [style=none] (39) at (-5.75, 0) {};
		\node [style=none] (40) at (-5.75, 1) {};
		\node [style=none] (41) at (-7.75, 0.5) {};
		\node [style=none] (42) at (-5.75, 0.5) {};
		\node [style=none] (43) at (-1.75, 1) {$\mu_i$};
		\node [style=none] (44) at (-6.75, 1) {$\mu_{i+1}$};
		\node [style=none] (45) at (-4.25, 0.5) {$\mu_{i+1} | \mu_i$};
		\node [style=rect] (46) at (1.25, -0.75) {$\xi_0$};
		\node [style=none] (47) at (0.25, 0.5) {};
		\node [style=none] (48) at (0.25, -2) {};
		\node [style=none] (49) at (-0.5, 0.5) {};
		\node [style=none] (50) at (0, 0.5) {};
		\node [style=none] (51) at (-0.5, -2) {};
		\node [style=none] (52) at (0, -2) {};
		\node [style=none] (53) at (-8, 0.5) {};
		\node [style=none] (54) at (-8.5, 0.5) {};
		\node [style=none] (55) at (-8.5, -2) {};
		\node [style=none] (56) at (-8, -2) {};
		\node [style=none] (57) at (-8.75, 0.5) {};
		\node [style=none] (58) at (-8.75, -2) {};
		\node [style=none] (59) at (6, 5) {};
		\node [style=none] (60) at (8, 5) {};
		\node [style=none] (61) at (6, 3) {};
		\node [style=none] (62) at (8, 3) {};
		\node [style=none] (63) at (10, 4) {};
		\node [style=none] (64) at (12, 4) {};
		\node [style=none] (65) at (10, 2) {};
		\node [style=none] (66) at (12, 2) {};
		\node [style=none] (67) at (7, 1) {};
		\node [style=none] (68) at (9, 1) {};
		\node [style=none] (69) at (7, -1) {};
		\node [style=none] (70) at (9, -1) {};
		\node [style=none] (71) at (11, 0) {};
		\node [style=none] (72) at (13, 0) {};
		\node [style=none] (73) at (11, -2) {};
		\node [style=none] (74) at (13, -2) {};
		\node [style=none] (75) at (5, 6) {};
		\node [style=none] (76) at (14, 6) {};
		\node [style=none] (77) at (14, -3) {};
		\node [style=none] (78) at (5, -3) {};
		\node [style=none] (79) at (5, 5.5) {};
		\node [style=none] (80) at (6, 4.5) {};
		\node [style=none] (81) at (6, 3.5) {};
		\node [style=none] (82) at (5, 2.5) {};
		\node [style=none] (83) at (8, 4.5) {};
		\node [style=none] (84) at (8, 3.5) {};
		\node [style=none] (85) at (10, 3.5) {};
		\node [style=none] (86) at (10, 2.75) {};
		\node [style=none] (87) at (10.5, 2) {};
		\node [style=none] (88) at (11.5, 2) {};
		\node [style=none] (89) at (12.5, 0) {};
		\node [style=none] (90) at (11.5, 0) {};
		\node [style=none] (91) at (11, -0.5) {};
		\node [style=none] (92) at (11, -1.25) {};
		\node [style=none] (93) at (9, 0.5) {};
		\node [style=none] (94) at (9, -0.5) {};
		\node [style=none] (95) at (8.5, -1) {};
		\node [style=none] (96) at (7.5, -1) {};
		\node [style=none] (97) at (6.75, -3) {};
		\node [style=none] (98) at (9.25, -3) {};
		\node [style=none] (99) at (4, 5.5) {};
		\node [style=none] (100) at (4, 2.5) {};
		\node [style=none] (101) at (6.75, -4) {};
		\node [style=none] (102) at (9.25, -4) {};
		\node [style=none] (103) at (7, 4) {};
		\node [style=none] (104) at (11, 3) {};
		\node [style=none] (105) at (8, 0) {$\mu_{i-1}$};
		\node [style=none] (106) at (12, -1) {$\xi_{i-1}$};
		\node [style=none] (107) at (7, 4) {$\xi_i$};
		\node [style=none] (108) at (11, 3) {$\mu_i$};
		\node [style=none] (109) at (4, 6) {$\mu_{i+1}$};
		\node [style=none] (110) at (4, 3) {$\mu_{i+1}$};
		\node [style=none] (111) at (7.5, -3.75) {$\xi_{i-1}$};
		\node [style=none] (112) at (10, -3.75) {$\xi_{i-1}$};
		\node [style=none] (113) at (-15, 5.75) {\Large \textbf{(a)}};
		\node [style=none] (114) at (-8.75, 1.75) {\Large \textbf{(b)}};
		\node [style=none] (115) at (2.25, 5.75) {\Large \textbf{(c)}};
	\end{pgfonlayer}
	\begin{pgfonlayer}{edgelayer}
		\draw [style=pointy1] (1.center) to (0);
		\draw [style=pointy1] (0) to (2);
		\draw (3.center) to (4.center);
		\draw (3.center) to (5.center);
		\draw (5.center) to (4.center);
		\draw [style=pointy1, bend right] (7.center) to (2);
		\draw [style=pointy1] (2) to (8.center);
		\draw (14.center) to (15.center);
		\draw (14.center) to (17.center);
		\draw (15.center) to (16.center);
		\draw (17.center) to (16.center);
		\draw [style=pointy1] (19.center) to (18.center);
		\draw [style=pointy1] (12.center) to (13.center);
		\draw (26.center) to (27.center);
		\draw (26.center) to (29.center);
		\draw (27.center) to (28.center);
		\draw (29.center) to (28.center);
		\draw [style=pointy1] (31.center) to (30.center);
		\draw [style=pointy1] (24.center) to (25.center);
		\draw (37.center) to (38.center);
		\draw (37.center) to (40.center);
		\draw (38.center) to (39.center);
		\draw (40.center) to (39.center);
		\draw [style=pointy1] (42.center) to (41.center);
		\draw [style=pointy1] (35.center) to (36.center);
		\draw [style=pointy1, bend right=45] (48.center) to (46);
		\draw [style=pointy1, bend right=45] (46) to (47.center);
		\draw [style=pointy1] (50.center) to (49.center);
		\draw [style=pointy1] (51.center) to (52.center);
		\draw [style=pointy1] (53.center) to (54.center);
		\draw [style=pointy1] (55.center) to (56.center);
		\draw [style=pointy1, bend left=270, looseness=1.25] (57.center) to (58.center);
		\draw (75.center) to (76.center);
		\draw (75.center) to (78.center);
		\draw (78.center) to (77.center);
		\draw (76.center) to (77.center);
		\draw (59.center) to (60.center);
		\draw (60.center) to (62.center);
		\draw (59.center) to (61.center);
		\draw (62.center) to (61.center);
		\draw (63.center) to (65.center);
		\draw (63.center) to (64.center);
		\draw (64.center) to (66.center);
		\draw (66.center) to (65.center);
		\draw (67.center) to (68.center);
		\draw (68.center) to (70.center);
		\draw (70.center) to (69.center);
		\draw (69.center) to (67.center);
		\draw (71.center) to (72.center);
		\draw (72.center) to (74.center);
		\draw (74.center) to (73.center);
		\draw (73.center) to (71.center);
		\draw [style=pointy1] (79.center) to (80.center);
		\draw [style=pointy1] (83.center) to (85.center);
		\draw [style=pointy1] (88.center) to (89.center);
		\draw [style=pointy1] (91.center) to (93.center);
		\draw [style=pointy1] (95.center) to (98.center);
		\draw [style=pointy1] (97.center) to (96.center);
		\draw [style=pointy1] (94.center) to (92.center);
		\draw [style=pointy1] (90.center) to (87.center);
		\draw [style=pointy1] (86.center) to (84.center);
		\draw [style=pointy1] (81.center) to (82.center);
		\draw [style=pointy1] (99.center) to (79.center);
		\draw [style=pointy1] (82.center) to (100.center);
		\draw [style=pointy1] (98.center) to (102.center);
		\draw [style=pointy1] (101.center) to (97.center);
	\end{pgfonlayer}
\end{tikzpicture}
  }
  \caption{\label{fig:circ} From $\Cat{BayesNets}$ to
    $\Cat{NeurCirc}$. \textbf{(a)} One level of a hierarchical generative
    model. $g_{i+1}$ is a transfer function and the noise source $z_i$ is
    standard normal. $\mu_{i+1}$ encodes, say, the latent banana state, while
    $\mu_i$ encodes the joint feature space. The right hand side summarizes the
    left as a single Markov kernel. \textbf{(b)} One layer of a predictive
    coding hierarchy. Downward signals encode the generative model, while upward
    signals encode the recognition density's approximate Bayesian inverse; at
    bottom is a perceptual error unit. \textbf{(c)} A $\Cat{DDS}$ object
    corresponding to (b). NB: Just as the outer box can be wired up to other
    layers in the cortical hierarchy, the opaquely labeled inner boxes can be
    filled in with further structure (\emph{eg.}, for within-layer hidden
    states). The advantage of this framework is that this `zooming in' can be
    performed in a completely rigorous manner.}
\end{figure}

Above, we moved rigorously from an abstract compositional concept model to a
probabilistic state-space representation. Here, we see how this could be mapped
onto neural circuits, emphasizing the compositionality of such `Bayesian brain'
circuits. Recall that under the Bayesian brain hypothesis, the cortical
hierarchy can be interpreted as a message-passing architecture solving an
approximate inference problem \citep{friston2012history}. For simplicity, we
restrict ourselves here to discrete-time models, and thus make the following
definition:

\defn{A \textbf{discrete dynamical system} (DDS) is a pair of functions $f^{upd}
  : I \times S \to S$ and $f^{out} : S \to O$, which we call \emph{update} and
  \emph{output}. $S$ is the \emph{state space}, $I$ the \emph{input space} and
  $O$ the \emph{output space}. There is a category $\Cat{DDS}$ whose objects are
  DDSs and whose arrows are wirings of outputs to inputs
  \citep{spivak2015steady,schultz2016dynamical}.}

Predictive coding models under the free energy principle have a somewhat
formidable reputation. The following theorem, whose proof we only sketch,
establishes that this is unjustified: compositionality ensures that complexity
can be reduced to its elements.

\thm{There is a functor $\mathcal{F} : \Cat{BayesNets} \to \Cat{NeurCirc}
  \xhookrightarrow{} \Cat{DDS}$, where $\Cat{NeurCirc}$ is a subcategory of
  $\Cat{DDS}$ consisting of neural circuits of the form illustrated
  below. $\mathcal{F}$ maps measurable spaces to layers of neurons, and
  conditional probability distributions to message-passing architectures.}

Let $\mu_i$ represent neural activity at cortical hierarchy level $i$. The key
idea is that the brain embodies both a \emph{generative} model $\mu_i |
\mu_{i+1}$ (\figref{fig:circ}(a))---\emph{e.g.}, modelling expected transitions
in latent states $\mu_{i+1}$ of the environment and how these states cause
perceptual signals $\mu_i$---and a \emph{recognition} model $\mu_{i+1} | \mu_i$:
an approximate Bayesian inversion of the generative model, taking perceptual
signals back to inferred latent causes. This is illustrated in
\figref{fig:circ}(b). At each stage, neural dynamics minimize the KL divergence
between the recognition and generative densities \citep{buckley2017free}. Under
some simplifying assumptions, there is an upper bound on this divergence called
the free energy \citep{buckley2017free}, written:
\begin{align}
  F = \sum_i \epsilon_i^T \Sigma_i^{-1} \epsilon_i + \ln | \Sigma_i | \, ; \label{eq:F} \quad
  \epsilon_i = \overleftarrow{\mu_i} - g_{i+1}(\overrightarrow{\mu_{i+1}}) \, .
\end{align}
The over-arrows are purely syntactic sugar to indicate whether a quantity is
transmitted from sense organs to higher cortex ($\gets$) or vice versa
($\to$). We call the $\mu_i$ \textbf{state units} and define \textbf{error
  units} as $\xi_i = \Sigma_i^{-1} \epsilon_i$. We construct the dynamics of the
circuits encoding $\mu_i$ to minimize $F$, and so we write:
\begin{align}
  \partial_{\mu_i} F &= -\partial_{\mu_i} g(\mu_i)^T \xi_{i-1} + \xi_i \, , \\
  \mu_i(t + 1)  = \Delta_i[\mu_i(t)] - \kappa \partial_{\mu_i} F
  &= \Delta_i[\mu_i(t)] - \kappa \left( \xi_i(t) - \partial_{\mu_i} g[\mu_i(t)]^T \xi_{i-1}(t) \right) \, ,
\end{align}
where $\Delta_i$ is the transition function of the generative dynamics, $g$ is
the inter-layer transfer function, and $\kappa$ is a learning rate. Note that
state units thus receive error signals from the current layer and that
immediately below, whereas error units receive state signals from the current
layer and that immediately above, as illustrated in
\figref{fig:circ}(c). Writing $\xi_i \in E_i$ and $\mu_i \in M_i$, and letting
$(E_i \times M_i) \subseteq S_i$, $I_i = (E_{i-1} \times M_{i+1})$, and $O_i =
(E_i \times M_i)$, we can package our definitions of $\mu_i(t)$ and $\xi_i(t)$
into the forms $f_i^{upd} : I_i \times S_i \to S_i$ and $f_i^{out} : S_i \to
O_i$ required by $\Cat{DDS}$.

To demonstrate the functoriality of $\mathcal{F}$, we particularly need to show
that it preserves composition. This is immediate from the associativity of the
sum in \eqref{eq:F} and compositionality of output-input wiring in $\Cat{DDS}$.

\section{... and back again} \label{sec:back}

Under the Bayesian brain hypothesis, much of neuroscience can be interpreted as
inferring the form of the generative model implied by the structure of neural
circuits; that is, as inverting the functor $\mathcal{F}$. When we have a
situation involving two functors $\mathcal{F} : \Cat{C} \to \Cat{D}$ and
$\mathcal{G} : \Cat{D} \to \Cat{C}$ such that the composites $\mathcal{F} \circ
\mathcal{G}$ and $\mathcal{G} \circ \mathcal{F}$ lead you back somehow to your
starting point, then we say that $\mathcal{F}$ and $\mathcal{G}$ are
\textbf{adjoint}.

This `modelling' adjunction is a largely syntactic one: the structure of a
probabilistic model is mapped to the structure of a neural circuit. Even so, it
is a deep one, similar to that between proofs and programs
\citep{baez2010physics}. Cognitive scientists are however typically interested
in the semantics of these circuits, \emph{i.e.}, their behavior. And indeed
there is a related adjunction between formal machines and their behaviors
\citep{jacobs2017introduction} that is already used in computer science for
verification \citep{cirstea2000algebra} and testing
\citep{pavlovic2006testing}. Yet another related adjunction is that which gave
rise to our view of distributions as convex spaces \citep{cho2015introduction}.

To illustrate this (see figure on accompanying poster): first, we need a
\emph{closed} dynamical system, which means supplying any unwired inputs with
some data, $1 \to I$, and discarding any unused output $O \to 1$. This gives a
closed system $S \to G(S)$. Given an initial state $s_0 : 1 \to S$, we can
generate behavior distributions by iterating the system. Necessarily, these live
in a convex space which includes the concept spaces we began with.

\section{Conclusions, in context} \label{sec:conc}

Categorical methods supply us with a compositional type theory for science:
different concepts and models are collected into categories, which tell us how
to compose when the types match. Categories themselves are collected into a
`meta'-category, which provides a context for translating ideas between
settings, thereby elucidating common structures; for instance, the
`symbolic-vs-connectionist' debate
\citep{fodor1988connectionism,smolensky1987constituent} is dissolved by
functorial semantics. This kind of rigorous conceptual modularity will be
increasingly important as we strive to build ever more complex yet interpretable
models, and understand ever more complex systems.

Space constraints mean much has been omitted: in particular, there is much more
to say on the compositional structure of interacting neural circuits and their
supervenient representations. There is also much to say about contextuality,
which can be seen as a kind of failure of compositionality. Even so, category
theory supplies the tools to deal with it rigorously, via fibrations and
sheaves; such a treatment once more sheds light on the neural
substrate. Finally, the treatment here has been theoretical: but simulations are
in progress.

\small

\printbibliography
\end{document}